# *Development of an Automated Web Application for Efficient Web Scraping: Design and Implementation*

Alok Dutta, Nilanjana Roy, Rhythm Sen, Sougata Dutta , Prabhat Das

*Department of Computer Science and Engineering, School of Engineering and Technology, Adamas University, Kolkata, India*

**Abstract:**
This paper presents the design and implementation of a user-friendly, automated web application that simplifies and optimizes the web scraping process for non-technical users. The application breaks down the complex task of web scraping into three main stages: fetching, extraction, and execution. In the fetching stage, the application accesses target websites using the HTTP protocol, leveraging the requests library to retrieve HTML content. The extraction stage utilizes powerful parsing libraries like BeautifulSoup and regular expressions to extract relevant data from the HTML. Finally, the execution stage structures the data into accessible formats, such as CSV, ensuring the scraped content is organized for easy use. To provide personalized and secure experiences, the application includes user registration and login functionalities, supported by MongoDB, which stores user data and scraping history. Deployed using the Flask framework, the tool offers a scalable, robust environment for web scraping. Users can easily input website URLs, define data extraction parameters, and download the data in a simplified format, without needing technical expertise. This automated tool not only enhances the efficiency of web scraping but also democratizes access to data extraction by empowering users of all technical levels to gather and manage data tailored to their needs. The methodology detailed in this paper represents a significant advancement in making web scraping tools accessible, efficient, and easy to use for a broader audience.

## Introduction:

Web scraping, also known as data harvesting or data crawling, has been a fundamental technique since the inception of the internet. While contemporary understanding often associates web scraping with the extraction of vast amounts of information from websites, its original purpose was to enhance the usability of the World Wide Web (WWW) [1]. The early WWW, though much less visual and smaller than today's internet, introduced three critical features that remain integral to web scraping tools: embedded hyperlinks for website navigation, Uniform Resource Locators (URLs) to assign scrapers to specific sources, and web pages containing various types of data, such as text, photos, videos, and audio files [1].

The journey of web scraping began with the creation of the first web browser by Tim Berners-Lee, the inventor of the World Wide Web. This browser was an HTTP web page running on a server on Berners-Lee's computer [1]. Following this, the first web robot, the World Wide Web Wanderer, emerged in 1993. Created by Matthew Gray at the Massachusetts Institute of Technology, the Wanderer was a Perl-based web crawler that measured the size of the WWW and later helped create an index known as the Wandex [2].

In the same year, the internet witnessed the launch of Jump Station, the first crawler-based web search engine. Developed by Jonathon Fletcher, a systems administrator at the University of Stirling in Scotland, Jump Station automated the indexing of millions of web pages, which was previously a manual, labor-intensive task (3). By indexing 275,000 entries across 1,500 servers, Jump Station democratized access to vast amounts of information and marked a significant advancement in web accessibility and information retrieval. It used headings and document titles to index web pages found through a linear search and displayed results as URLs matching users' keywords, much like the modern Google Search [3].

As the internet evolved, so did the tools for web scraping. Beautiful Soup, a Python library, emerged as a popular tool among programmers for parsing HTML and XML documents. This library simplified the process of understanding site structures and extracting content from HTML containers, thereby saving programmers considerable time and effort [4]. Beautiful Soup's capabilities made it one of the most sophisticated libraries for web scraping, enabling users to efficiently extract text, pictures, and other information from websites [4].

The rise of visual web scrapers further transformed the landscape of web scraping. Companies developed user-friendly visual web scraping software platforms that allowed users to manually highlight the information they wanted to extract into formats like Excel spreadsheets or databases. These platforms, designed with intuitive interfaces, catered to non-programmers and simplified the data extraction process (5). Users could select elements for extraction, define the extraction sequence, and initiate the process with a simple click, allowing the visual web scraper to automatically populate the designated spreadsheets or databases with the scraped data [5].

As the internet continues to burgeon with vast amounts of information, web scraping remains a crucial method for collecting, analyzing, and utilizing data for various purposes, including research, business intelligence, and content aggregation [6]. This paper explores the evolution of web scraping tools and techniques, the implications for data privacy and security, and the best practices for responsible data handling. It also provides a detailed guide on using Python-based tools for web scraping, focusing on practical implementation and real-world applications.

## Applications and Types of Web Scraping Tools:

Web scraping has numerous applications across various industries, including insurance, banking, finance, trading, e-commerce, sports, and digital marketing. It plays a crucial role in decision-making, lead generation, sales, risk management, strategy formulation, and the development of new products and services such as price intelligence, brand monitoring, business automation, real estate insights, and content marketing [7].

Different types of web scraping tools are designed to meet diverse needs and skill levels:

- Web Scraper Browser Extension: This Chrome extension allows users to create a sitemap that specifies how to navigate a website and what data to extract [8].
- Self-Built Web Scrapers: Custom-built scrapers tailored to specific requirements, which necessitate advanced programming skills, especially in Python [9].

- Cloud Web Scrapers: Operating on remote servers provided by service providers, these scrapers avoid burdening the user's computer and typically offer advanced features, though they can be costly [10].
- Web Scraping Software: These tools include sophisticated features like IP rotation, JavaScript execution, proxy management, and anti-bot measures, making them suitable for complex scraping tasks [11].
- User Interface Web Scrapers: Featuring advanced user interfaces that render entire websites, these scrapers allow users to click on the desired data, making them accessible even to non-programmers [12].

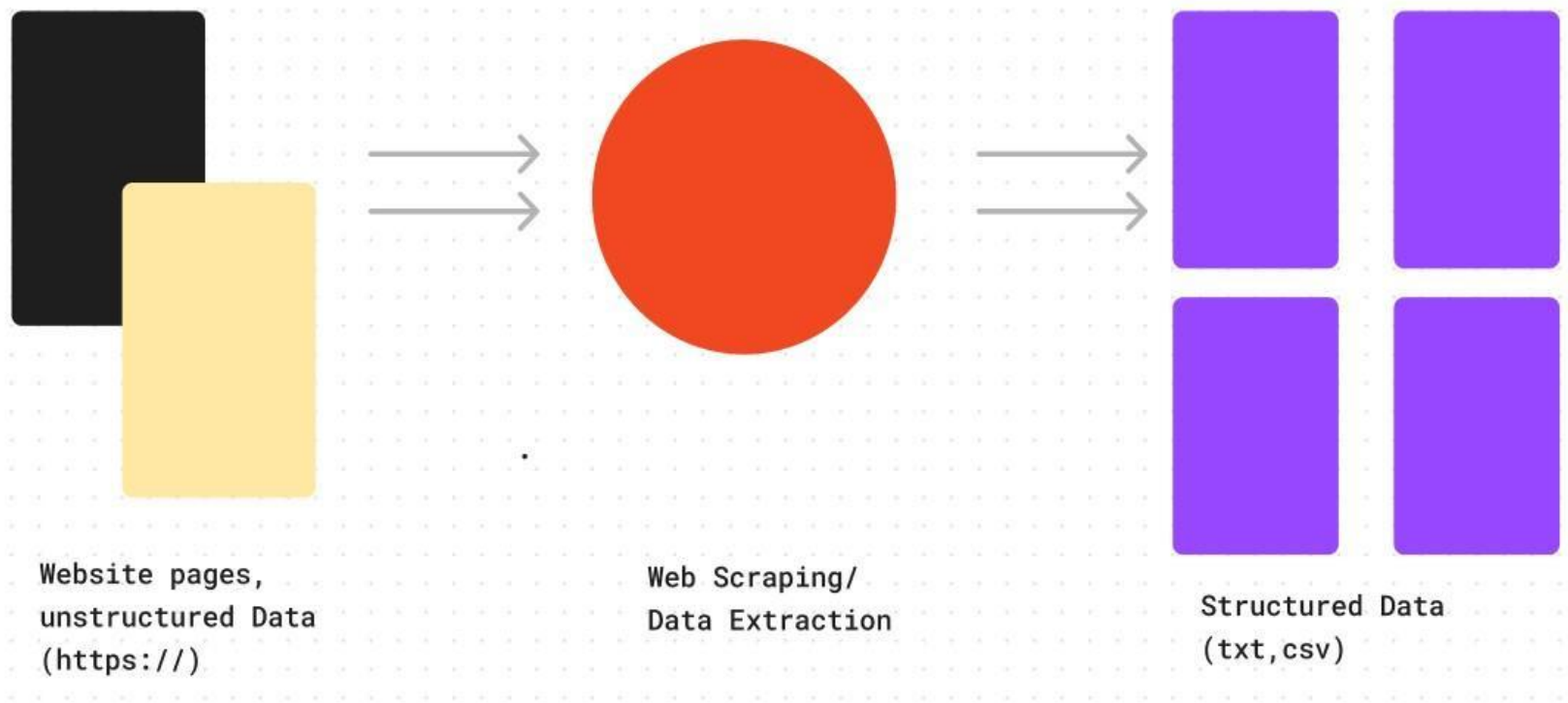

**Fig. 1.** Data Extraction From Websites [6]

Despite the utility of web scraping, non-technical users often struggle with the intricacies involved in the process. The necessity to understand and write code, manage data extraction logic, and handle data formatting tasks makes web scraping inaccessible to many potential users. Furthermore, the lack of user-friendly tools exacerbates this issue, leaving a gap in the market for solutions that democratize access to web scraping capabilities. As a result, there is a pressing need for an automated and user-friendly web application that simplifies the entire web scraping process, making it accessible to individuals regardless of their technical proficiency.

This research aims to design and implement an automated web application that optimizes the web scraping process. The proposed solution will streamline the complex steps of web scraping into a user-friendly interface, enabling users to fetch, extract, and execute data scraping tasks without requiring in-depth technical knowledge. By addressing the existing problem, this application seeks to empower users to efficiently gather and manage data tailored to their specific needs, thereby enhancing the overall effectiveness of web scraping.

**Literature Review:**

Web scraping has become an essential tool for extracting data from websites, enabling researchers and businesses to gather valuable information for various purposes. Several studies have explored different methods and techniques for web scraping, focusing on their effectiveness, efficiency, and limitations.

A study conducted at the General Sir John Kotelawala Defence University in Sri Lanka (2020) compared various methods of data extraction from websites. The study aimed to identify the most suitable approach for specific data extraction tasks based on factors such as accuracy and scalability. The findings of this study provide valuable insights into the optimal use of web scraping techniques for extracting data from websites [13].

In a study by Satyan Mundke and Sandeep Shreekumar (2022), the importance of web scraping in e-commerce businesses was highlighted. The authors emphasized that web scraping is essential for gathering data such as product details, pricing, and competitor insights. This data enables e-commerce businesses to monitor market trends, optimize pricing strategies, automate tasks, and enhance their offerings, ultimately leading to better decision-making and competitiveness in the market [14].

Justin Yek's guide on "How to scrape websites with Python and BeautifulSoup" (2020) provides a practical overview of web scraping using Python and BeautifulSoup. The guide explains how Python code can be used along with the BeautifulSoup library to extract data from web pages. By parsing the HTML structure of a webpage and extracting specific elements like text or links, BeautifulSoup facilitates the automation of information extraction from multiple web pages. This guide serves as a valuable resource for researchers and practitioners interested in utilizing web scraping for data collection and analysis [15].

In their work, Gustavo Pérez Molano and Jeffrey Duffany, Ph.D., provide an overview of web scraping techniques, focusing on technical aspects and practical exercises. The study covers essential concepts and tools like Python and BeautifulSoup, offering readers hands-on exercises to reinforce their understanding and skills in web scraping [16].

Kaajal Sharma and Gautam M Borkar conducted a comparative analysis of dynamic web scraping strategies to improve data acquisition efficiency. The study evaluates techniques for extracting data from dynamic web pages, aiming to identify the most effective approaches for capturing dynamic content [17].

Eloisa Vargiu and Mirko Urru explore the use of web scraping alongside collaborative filtering in web advertising. By extracting data from web sources to personalize online ads based on user behavior and preferences, this approach enhances ad targeting and improves the effectiveness of online advertising campaigns [18].

Harshit Nigam and Prantik Biswas provide a comprehensive guide to web scraping, covering tools, legal considerations, and practical Python implementation. This resource offers insights into the technical aspects, legal implications, and practical execution of web scraping tasks using Python [19].

Vidhi Singrodia, Anirban Mitra, and Subrata Paul discuss how web scraping extracts valuable data from online sources for purposes like market research and competitive analysis. The review may also address challenges and ethical considerations associated with web scraping practices [20].

Saram Han and Christopher K. Anderson examine how web scraping is used in the hospitality industry. Their study provides insights into gathering and analyzing data, highlighting its potential for market analysis and understanding consumer behavior. The study also explores the benefits and challenges of web scraping in hospitality research [21].

Kunal Mehta, Maya Salvi, Rumil Dand, Vineet Makharia, and Prachi Natu delve into adaptive web scraping techniques aimed at handling dynamic website content. The review evaluates adaptive strategies such as dynamic DOM traversal, machine learning-based classification, browser automation, and rule-based scraping with heuristics. While these approaches offer potential for efficiently extracting data from dynamic pages, further research is necessary to refine and improve their effectiveness [22].

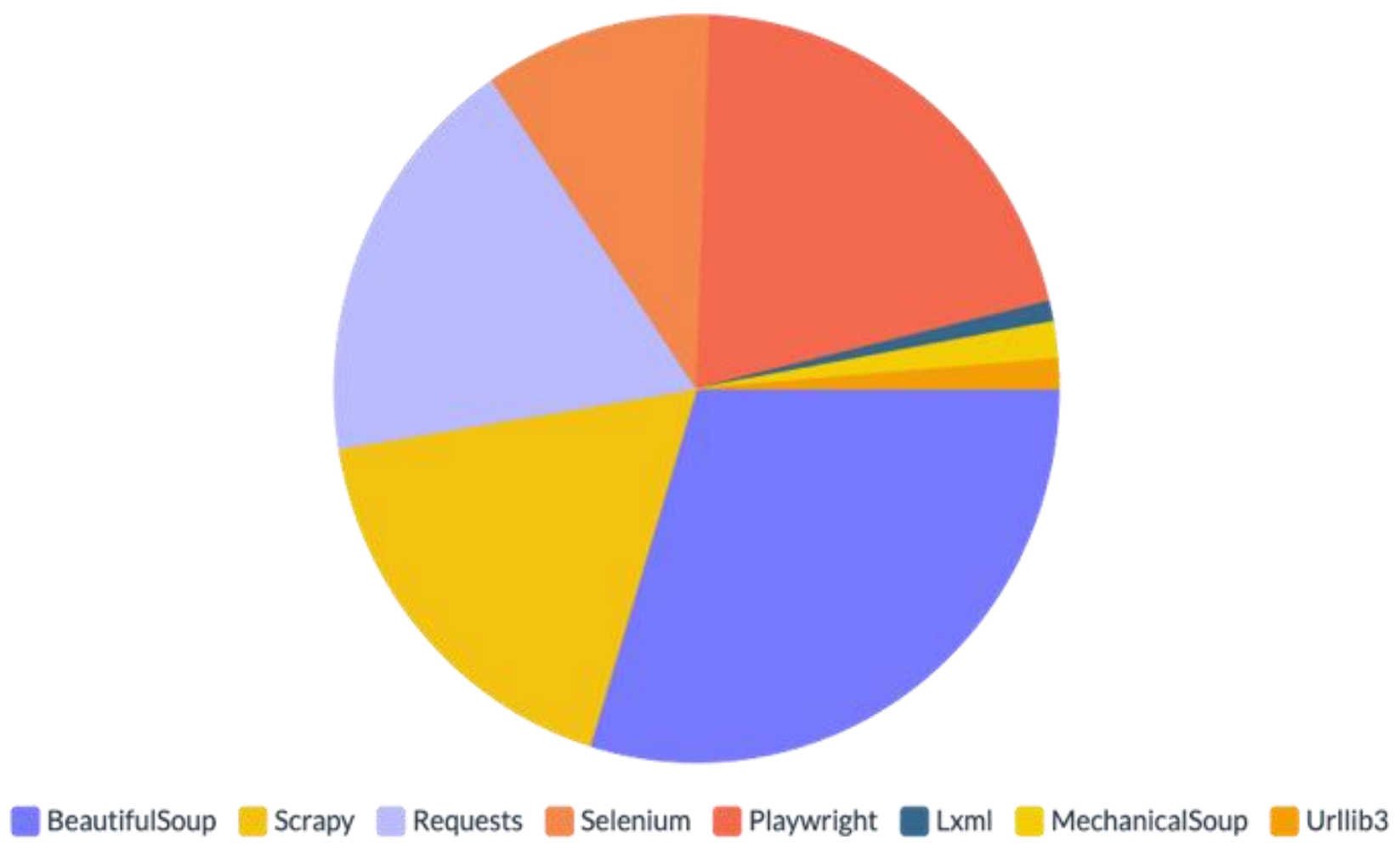

**Fig. 2.** Popularity of web scraping libraries in python [23].

**Table. 1.** Comparative Analysis of Python Web Scrapping Libraries

| Feature/Aspect | BeautifulSoup | Scrapy | Selenium | Requests | Urllib3 | lxml | MechanicalSoup |
|---|---|---|---|---|---|---|---|
| **Primary Use** | HTML/XML parsing | Web scraping framework | Browser automation | HTTP requests | HTTP requests | XML/HTML parsing | Automated web browsing |
| **Ease of Use** | Easy to learn, intuitive | Steeper learning curve | Moderate, requires Selenium WebDriver | Easy, simple API | Moderate | Requires understanding of XPath/ETree | Easy to use, based on Requests and BeautifulSoup |
| **Parsing Capabilities** | Strong for HTML/XML | Strong, with built-in tools | Limited, relies on other parsers | None, used for requests | None, used for requests | Very strong, high-performance | Moderate, uses BeautifulSoup |
| **Performance** | Moderate | High, suitable for large-scale scraping | Low to moderate, depends on browser | High | High | High | Moderate |
| **Dynamic Content Handling** | No | No | Yes, can handle JavaScript | No | No | No | No |
| **Headless Browser Support** | No | No | Yes | No | No | No | No |
| **Learning Curve** | Low | High | Moderate | Low | Low | Moderate | Low |
| **Concurrency Support** | No, requires external libraries | Yes, built-in support | Limited, dependent on WebDriver | No, requires external libraries | No | No | No |
| **Best For** | Small to medium-sized projects | Large-scale, complex scraping projects | Scraping sites with dynamic content | Simple data fetching tasks | Simple data fetching tasks | Large documents, performance-critical tasks | Form submissions, simple browsing |
| **Community & Documentation** | Strong | Very strong | Strong | Strong | Moderate | Strong | Moderate |
| **Installation** | Simple, one package | Requires several components | Requires WebDriver installation | Simple, one package | Simple, one package | Requires dependencies like libxml2 | Simple, one package |
| **License** | MIT License | BSD License | Apache License 2.0 | Apache License 2.0 | MIT License | BSD License | MIT License |

This study examined the performance of various web scraping libraries in terms of average memory consumption and run time. Based on our analysis:

- **lxml** consistently demonstrated the best performance, with the lowest memory consumption (10-30 MB) and the fastest average run time (917.67 ms), making it an ideal choice for efficient and lightweight parsing tasks.
- **BeautifulSoup** provided a good balance between memory consumption (20-50 MB) and average run time (1899 ms), making it a versatile choice for smaller to medium-scale scraping tasks.
- **Scrapy** performed well with a moderate memory footprint (50-150 MB) and a relatively fast run time (1365.67 ms), especially for larger-scale, asynchronous web scraping projects.
- **MechanicalSoup** offered moderate memory consumption (30-70 MB) and average speed (2019.67 ms), but it is slower compared to libraries like **lxml** and **Scrapy**.
- **Selenium** had the highest memory consumption (100-300 MB) and the slowest average run time (15397.33 ms), due to its browser automation capabilities. While it is more resource-intensive, Selenium remains essential for tasks requiring dynamic interaction with websites, such as handling JavaScript-heavy content.

The evolution of automation and intelligent data processing has established a strong foundation for developing modern web scraping systems that can adapt to changing web structures and heterogeneous data sources. Earlier research has focused on intelligent decision-making and dynamic data management using adaptive computational models such as fuzzy logic and sensor optimization techniques [24]–[28]. These works illustrate how uncertain and time-varying data environments can be handled through context-aware, rule-based processing an approach conceptually parallel to web scraping, where target websites often exhibit unpredictable structures, inconsistent markup, and variable response patterns. Fuzzy logic's ability to interpret ambiguous or incomplete information thus provides inspiration for designing robust extraction algorithms capable of maintaining accuracy under noisy or evolving HTML structures. Building on such intelligent paradigms, later studies emphasized scalability, clustering, and secure data management to efficiently process large-scale heterogeneous datasets [29]–[32]. These contributions highlight how AI-driven clustering and pattern-recognition frameworks can optimize post-scraping data organization, while secure communication and mobility management schemes ensure reliable data flow principles that are directly applicable to the design of an automated, intelligent web scraping framework that adapts, learns, and performs efficiently in dynamic online environments.

The literature review collectively highlights the diverse aspects of web scraping, including techniques, tools, applications, and challenges. The above survey provides a valuable insights for a researchers and practitioners interested in utilizing web scraping for data extraction and analysis. By understanding the trade-offs between memory usage and execution speed, users can make informed decisions about which library best suits their specific needs based on their performance requirements and the complexity of their scraping tasks.

**Methodology:**

Building upon the insights from the literature survey, the methodology in this research focuses on developing an efficient web scraping tool that simplifies the process into three core stages: fetching, extraction, and execution. The developed **Automated Web Scraper** is designed to balance performance (in terms of runtime and memory consumption) while also ensuring extended functionality such as data persistence in a database.

1. Fetching Stage:
   - The desired website containing the relevant information is accessed using the HTTP protocol.
   - The requests library is used to send an HTTP GET request to the target URL, retrieving the HTML page as a response.
   - Optimization Consideration: To improve the performance, libraries such as lxml are considered, as they have been shown to have minimal runtime and memory overhead.

2. Extraction Stage:
   - After retrieving the HTML page, the important data is extracted using HTML parsing libraries such as **BeautifulSoup** or **lxml**, depending on the complexity of the structure.
   - For more dynamic content, the **Selenium** library can be employed to interact with JavaScript-rendered components of the webpage. However, this is done cautiously since it incurs higher runtime and memory usage.
   - Regular expressions may also be used in tandem with parsing libraries to extract specific patterns of data.

3. Execution Stage:
   - Once the relevant data is extracted, it is converted into a structured format, typically a CSV file, for further analysis or presentation.

- In addition to saving the data to a CSV file, this tool integrates MongoDB for data persistence, allowing users to maintain a history of the data fetched, which can be critical for long-term analysis or audit purposes.
- Performance Trade-Off**:** Although persisting data to a database increases memory consumption, this feature adds value by enabling historical analysis, making the tool more suitable for extensive data collection workflows.

**Mathematical model of web-based web scrapping application:**

Graph Representation of the Webpage:

Let $G = (V, E)$ represent the web page, where:

- $V$ is the set of nodes, each corresponding to an element in the HTML DOM tree (such as `<div>`, `<p>`, `<table>`, etc.).
- $E$ is the set of directed edges that define the relationships between elements. For example, if node $v_i$ is the parent of node $v_j$, then there's an edge from $v_i$ to $v_j$.

Data Extraction Process: The process of extracting relevant content from this graph is modelled as a search problem:

- Initial Node Selection: Start with a root node $v_r$ (the HTML root or a predefined target container such as `<body>`).
- Traversal: By applying a traversal algorithm (such as Depth-First Search or Breadth-First Search) to explore the tree structure, looking for relevant content based on attributes, classes, or tags.
- Let $T$ be the set of all tags of interest (e.g., $T$ might contain tags like `<table>`, `<div>`, `<span>`).
- At each node $v$, check if the tag $t(v) \in T$. If yes, extract the data contained within that node.

Content Filtering: Once the data is extracted, filtering can be applied by using specific rules or regular expressions $R$:

- Defining a filter function $F(v)$ that extracts data based on rules $R$. For instance, $R$ might specify certain patterns in text (e.g., numeric data, dates, or keywords).

The filtered content from node $v$ is given by $F(v)$.

Data Structuring: The extracted data is structured into a usable format. If $D$ represents the set of all extracted data points, then the structuring process is a mapping function:

$$\phi : D \rightarrow \text{Structured Format (e.g., CSV, JSON)}$$

Where each data point in $D$ is mapped into a row or field in the structured output.

Execution Model: Let the total number of nodes traversed in the scraping process be $n$, and the number of relevant nodes containing extractable data be $m$. The efficiency of the scraping process can then be described as:

$$\text{Efficiency} = \frac{m}{n}$$

This provides a measure of how efficiently the scraping algorithm filters and extracts relevant content from the web page.

Formal Equation: Summing up, a web scraping process $S$ over a webpage $G = (V, E)$ can be modeled as:

$$S(G) = \phi\left(\sum_{v \in V} F\,(v)\right)$$

Where:

- $F(v)$ is the filtering function that selects relevant content from node $v$,
- $\phi$ structures the selected content into the desired output format.

**Pseudo Code of the automated web based scrapping application:**

```
function InitializeApp
    Setup Flask session
    Connect to MongoDB
end function
function ScrapePage(url)
    headers ← default headers
    r ← GetRequest(url, headers)
    r.raise_for_status()
    return ParseHTML(r.content)
end function
function GetData(content)
    class_contents ← {}
    elements ← FindAllWithClass(content)
    for all element in elements do
        class_name ← JoinClass(element)
        tagname ← element.name
        UpdateClassContents(class_contents, class_name, tagname, element.text)
    end for
    return class_contents
end function
function SaveToCSV(data)
    filename ← GenerateFilename(session["user_name"])
    OpenFile(filename, 'w') as csvfile
    writer ← csv.writer(csvfile)
    writer.writerow(['Class', 'Tag', 'Content'])
    for all class_name, class_data in data.items() do
        for all tag_name, contents in class_data['subclasses'].items() do
            for all content in contents do
                writer.writerow([class_name, tag_name, content])
            end for
        end for
    end for
    return filename
end function
function RouteHomepage
    return Render('login.html')
end function
function RouteAuth
    username, password ← GetFormData(request)
    auth_status ← CountDocuments(collection, {"username":username,"password":password})
    if auth_status = 1 then
        session["user_name"] ← username
        return Render("search_url.html")
    else
        return Render("login.html", msg="Invalid Credentials !")
    end if
end function
function RouteRegistration
    return Render("register.html")
end function
function RouteCreateAccount
    username, password ← GetFormData(request)
    exist_status ← CountDocuments(collection, {"username": username})
    if exist_status ≥ 1 then
        return Render("register.html", msg="Username Exist !")
    else
        InsertDocument(collection, {"username":username,"password":password})
```

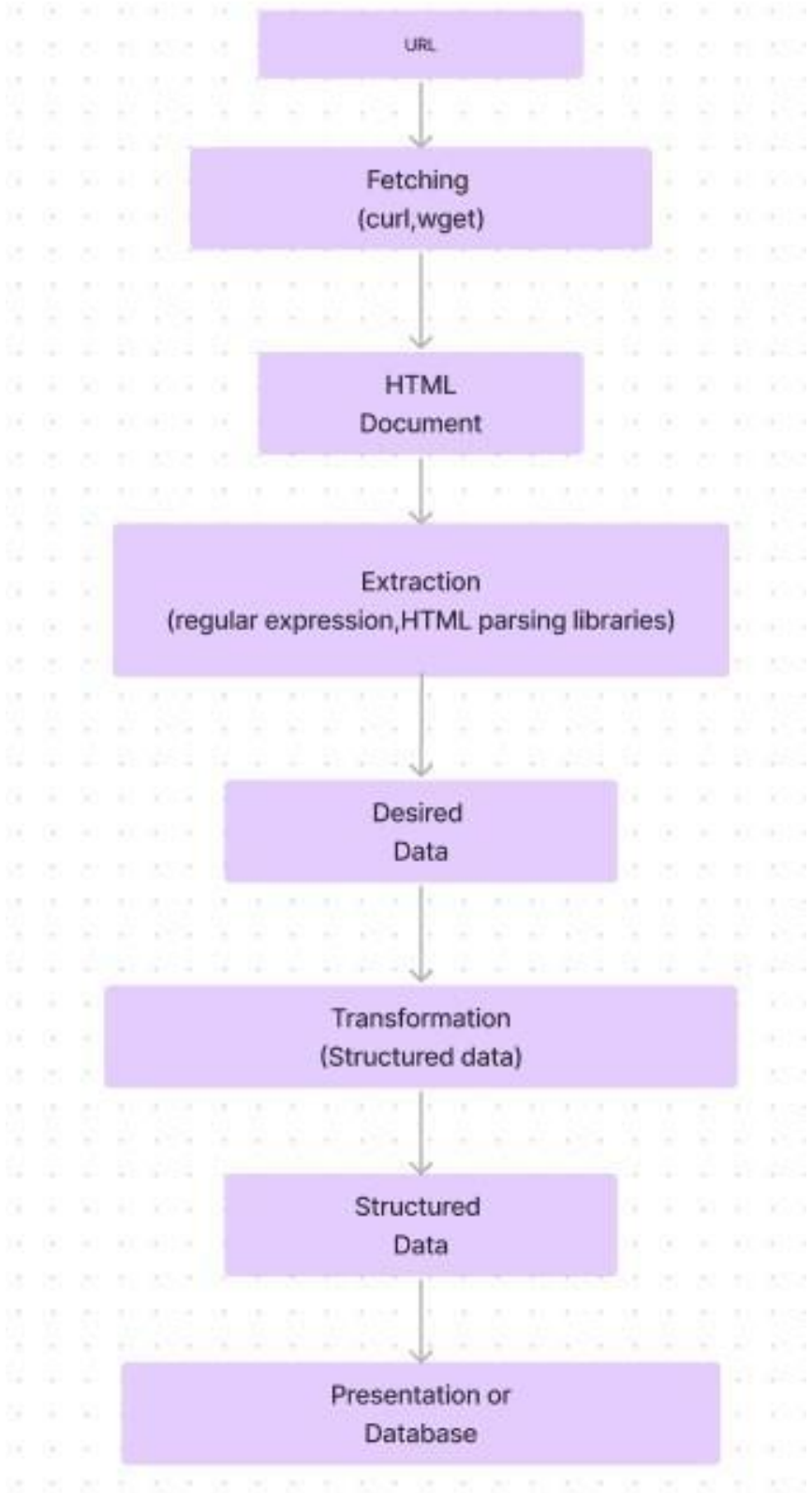

**Fig. 5.** Workflow of web-scraping tool

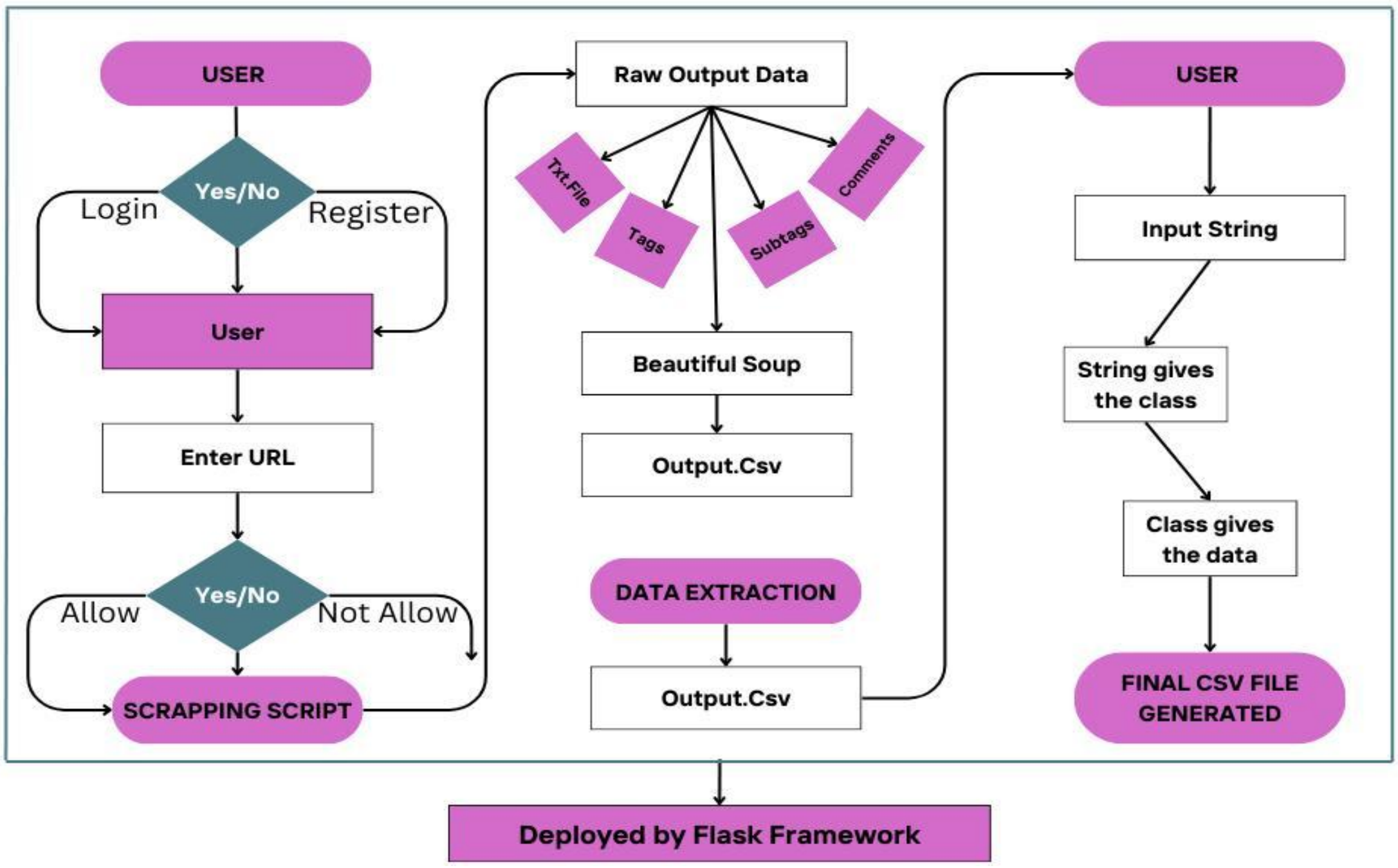

**Fig. 6.** System Architecture of web-scraping tool

The diagram shows a web scraping flowchart. The user starts by entering a URL. The program then checks to see if the user is registered. If not, the user is prompted to register. If the user is registered, the program then checks to see if the user wants to allow or not allow scraping. If the user allows scraping, the program then proceeds to extract the data from the website. The data is then saved to a CSV file. The CSV file is then deployed by the Flask Framework.

The methodology for the web scraping project involves the following steps, as demonstrated in the provided flowchart and case study:

1. User Registration and Login:
    - Users must register for an account on the website to access the scraping functionality.
    - Upon registration, users can log in to the website using their credentials.

2. Data Storage:
    - User data and scraping history are stored in MongoDB in separate files: one for user data and another for user history.

3. Scraping Process:
    - Once logged in, users can enter the URL of the website they want to scrape.
    - The program checks if the user has allowed scraping. If allowed, the program proceeds to extract the data from the website.

4. Data Extraction:
    - The extracted data is saved to a CSV file, which can be downloaded by the user.
    - To extract specific data, users can enter a string from the website to specify the information they want.

5. Data Refinement:
    - Users are provided with classes and subclasses connected to the entered string.
    - Users can select and copy a class, then paste it into the search bar for further refinement.

6. Final Data Output:
    - The final CSV file containing the refined data is generated for the user.

7. History Tracking:
    - Users can view their scraping history to see which URLs have already been scraped.

This methodology ensures that users can efficiently and effectively scrape data from websites for their specific needs, using the Flask framework for deployment and MongoDB for data storage.

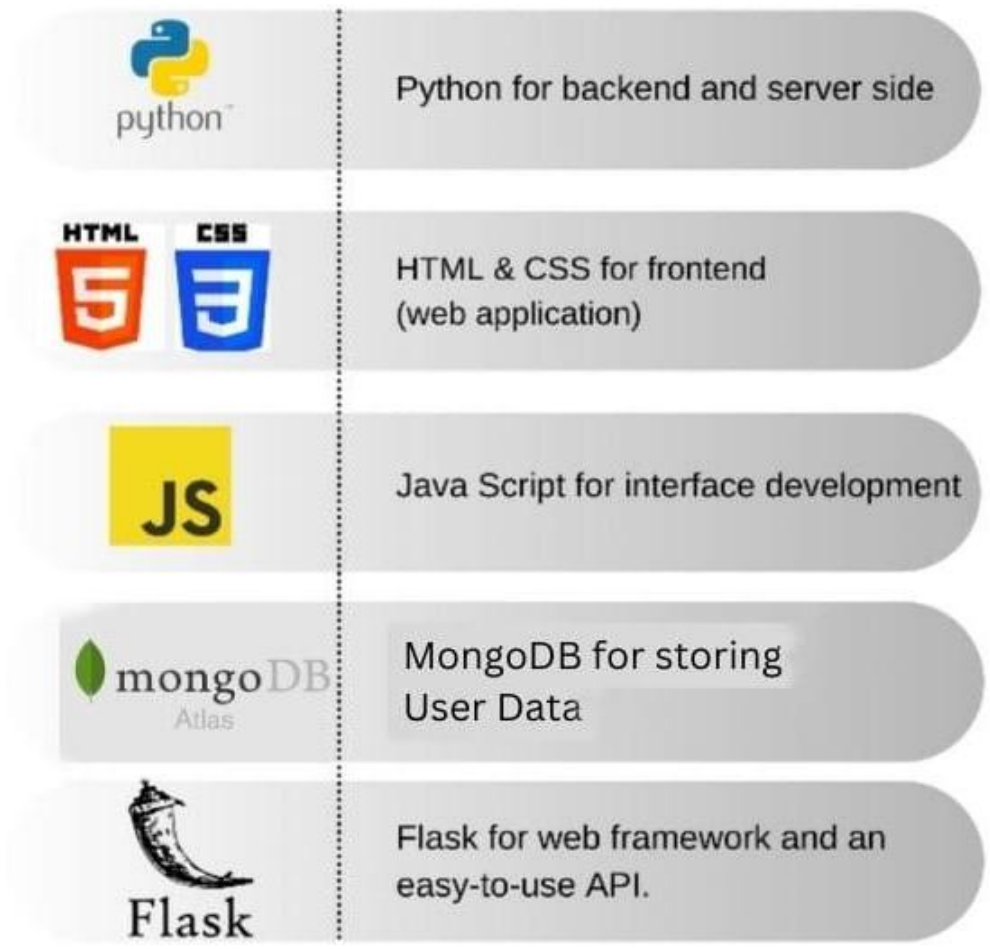

**Fig. 7.** Technology Stack of web-scraping tool

**Results and Discussions:**

The automated web scraping application developed using Python offers a comprehensive solution that combines robust data extraction tools with efficient web development capabilities. Python's libraries, particularly Beautiful Soup, play a crucial role in parsing HTML content and extracting relevant data from websites. Additionally, Flask, a lightweight web framework, provides a flexible environment for creating web applications with minimal overhead, making it an ideal choice for integrating with web scraping scripts.

The integration of these technologies enables the creation of a seamless workflow where users can input URLs or parameters through a Flask interface, initiating the web scraping process to collect targeted data. Subsequently, the scraped data is structured and saved into a CSV file, providing users with easily accessible and analysable data.
One of the key advantages of utilizing Python and Flask for web scraping automation is the scalability and customization options they offer. Flask's modular structure allows developers to seamlessly integrate features such as authentication, data validation, and error handling into the application, thereby enhancing its reliability and user experience.
From a practical standpoint, this automated web scraping application has significant potential across various domains, including e-commerce websites, competitive analysis, and data-driven decision-making. Its ability to efficiently extract and organize data makes it a valuable tool for businesses and researchers alike.

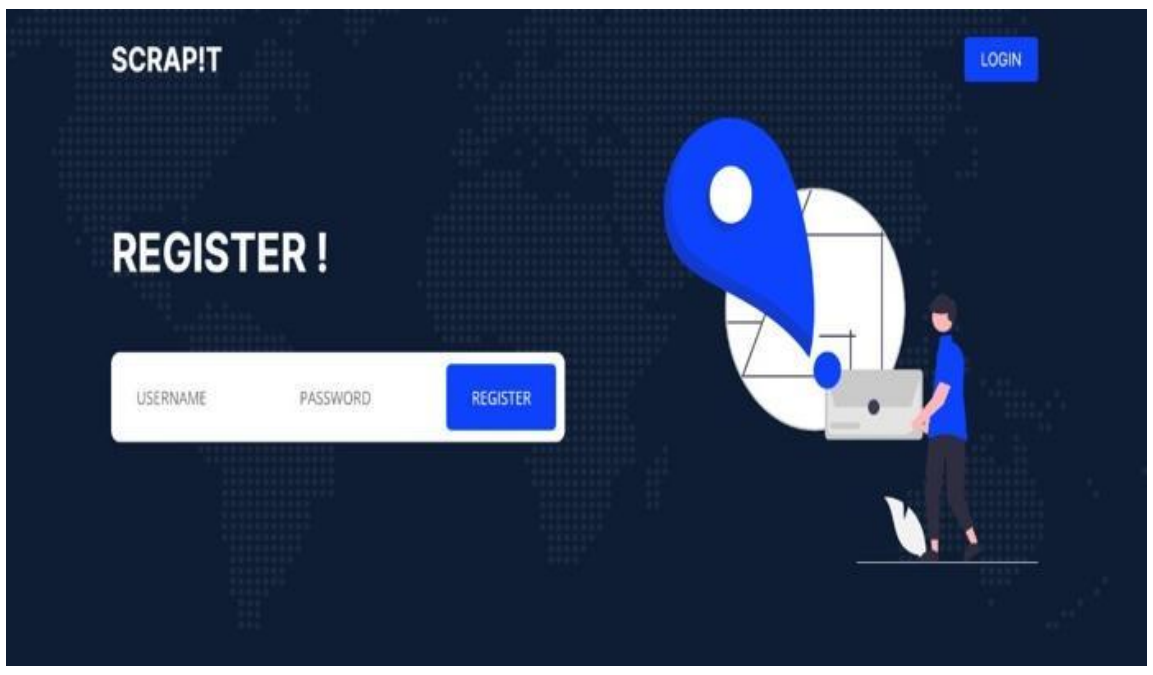

**Fig. 8.** User Interface of registration page

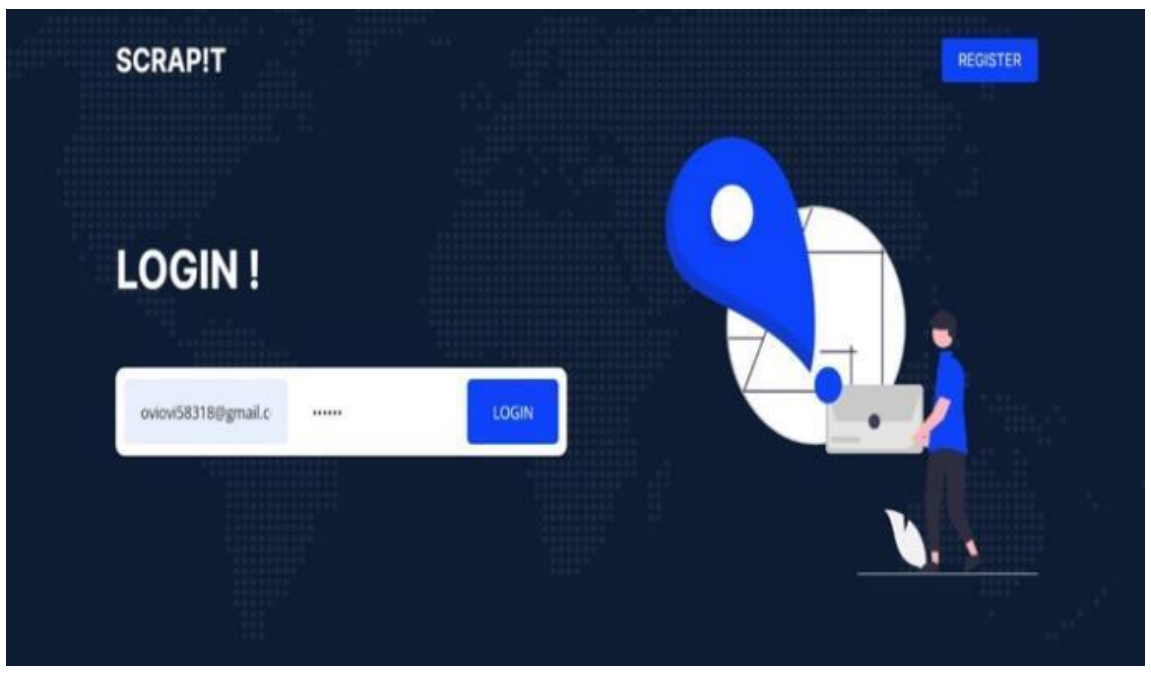

**Fig. 9.** User Interface of log in page

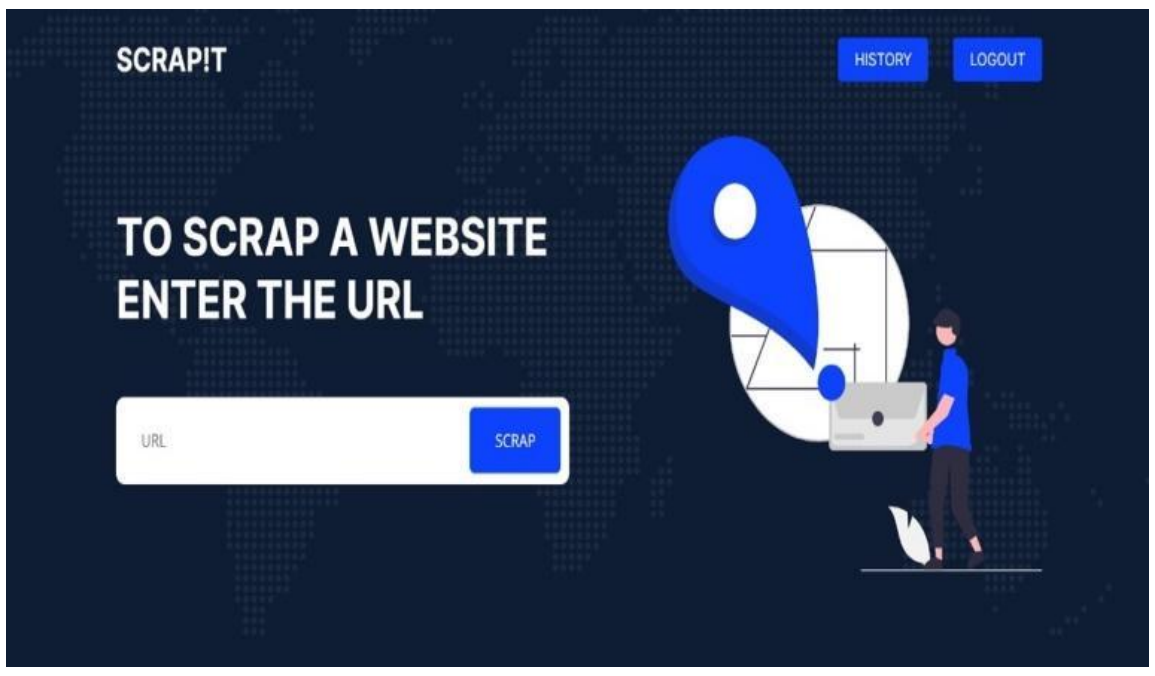

**Fig. 10.** User Interface of URL input page

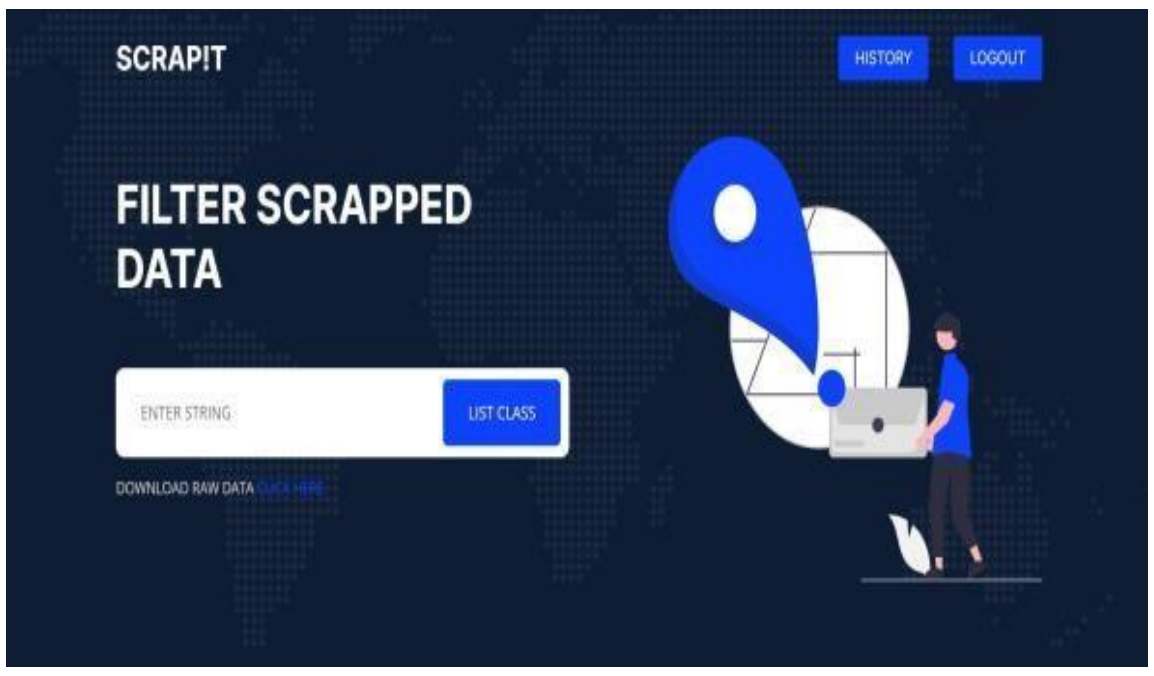

**Fig. 11.** User Interface of raw data download page

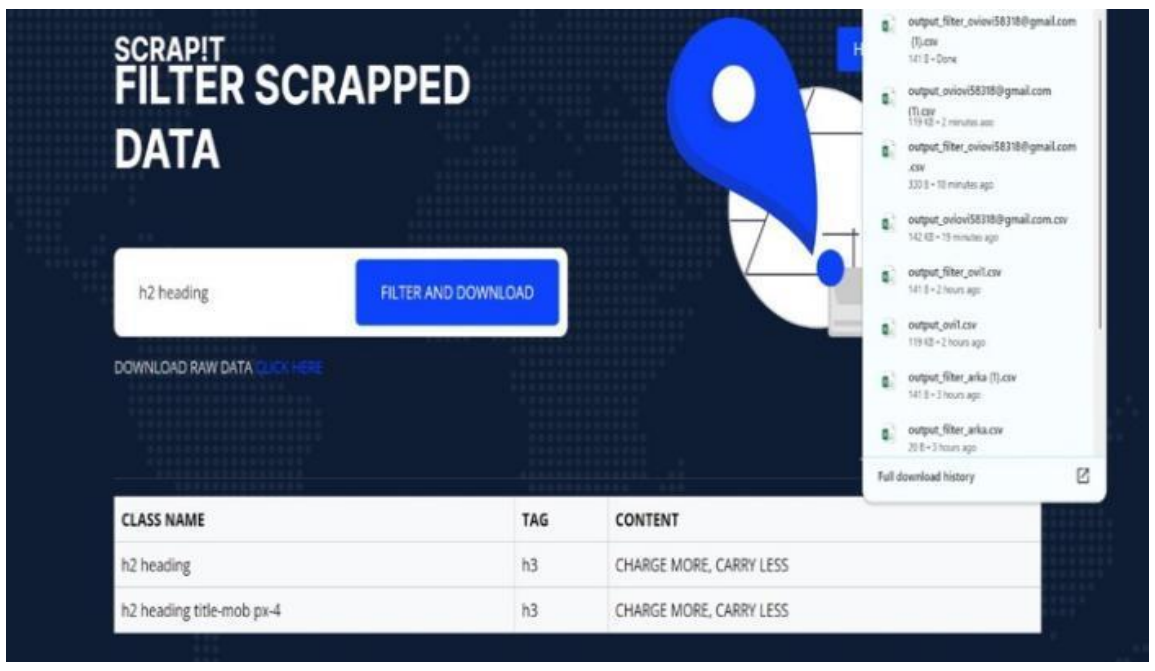

**Fig. 12.** User Interface of filtering scrapped data

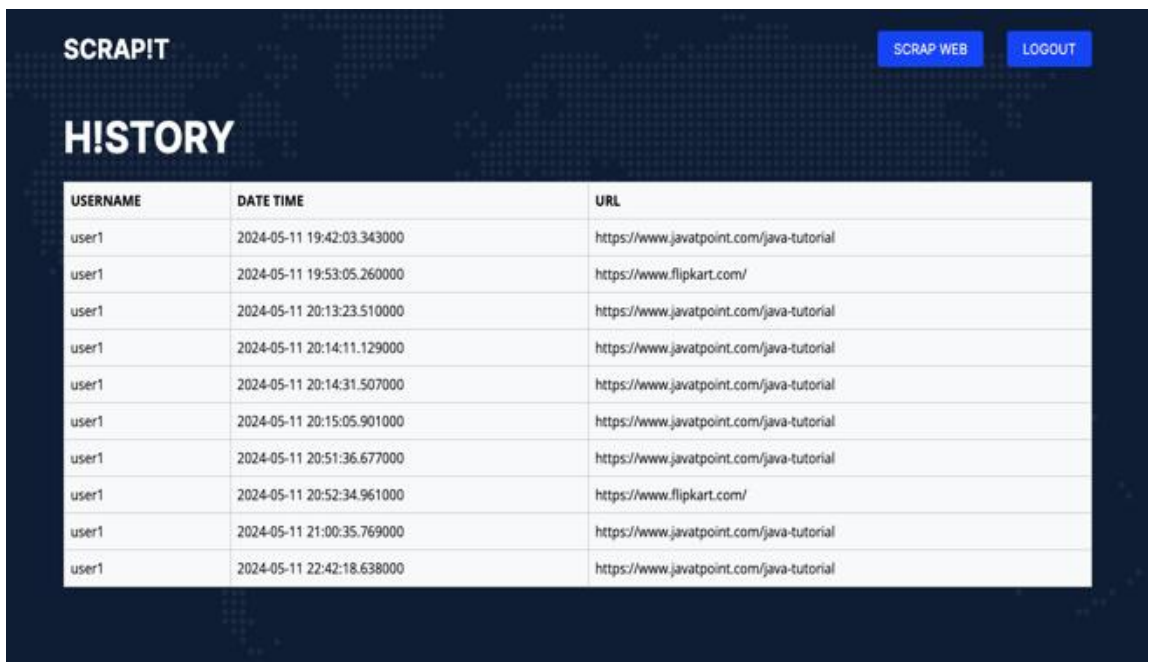

**Fig. 13.** User Interface of scrap history

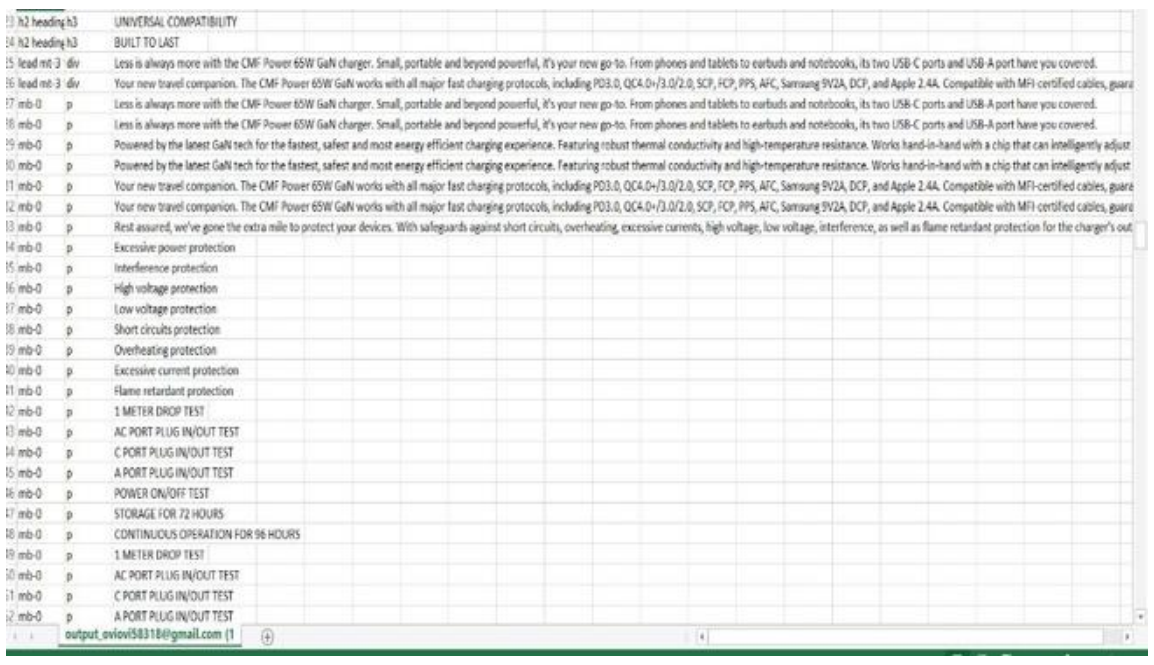

**Fig. 14.** Raw CSV data

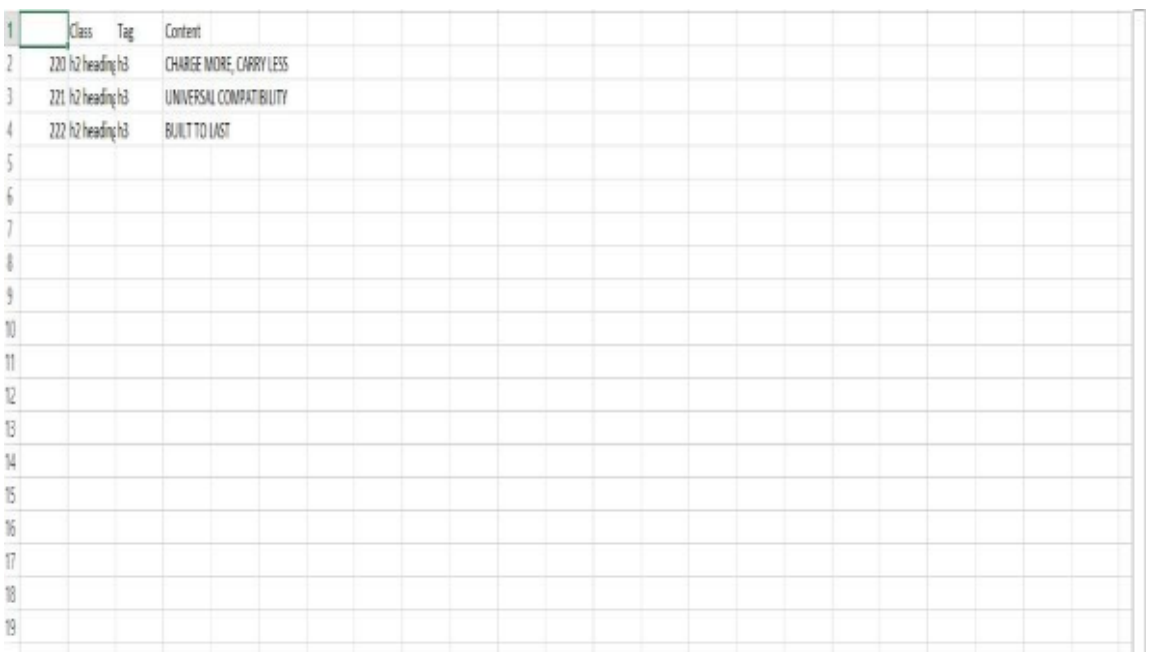

**Fig. 15.** Filtered CSV data

This section details the findings of the website scraping feasibility testing conducted on 400 websites across 20 categories: e-commerce, tutorial, static informative, portfolio, blogs, news, social media, forums, video sharing, photo sharing, online magazines, educational, real estate, government, health and fitness, technology reviews, travel and tourism, food and recipes, entertainment, and online directories.

**Table. 2.** Web scrapping tool results

| Website Category | Scrapable Websites | Websites Not Scrapable | Scrapability Rate |
|---|---|---|---|
| **E-commerce** | 20 | 5 | 80.00% |
| **Tutorial** | 22 | 3 | 88.00% |
| **Static Informative** | 25 | 0 | 100.00% |
| **Portfolio** | 24 | 1 | 96.00% |
| **Blogs** | 18 | 7 | 72.00% |
| **News** | 15 | 10 | 60.00% |
| **Social Media** | 10 | 15 | 40.00% |
| **Forums** | 14 | 6 | 70.00% |
| **Video Sharing** | 19 | 4 | 82.61% |
| **Photo Sharing** | 21 | 2 | 91.30% |
| **Online Magazines** | 20 | 5 | 80.00% |
| **Educational** | 23 | 2 | 92.00% |
| **Real Estate** | 17 | 8 | 68.00% |
| **Government** | 16 | 9 | 64.00% |
| **Health and Fitness** | 22 | 3 | 88.00% |
| **Technology Reviews** | 20 | 5 | 80.00% |
| **Travel and Tourism** | 18 | 7 | 72.00% |
| **Food and Recipes** | 24 | 1 | 96.00% |
| **Entertainment** | 23 | 2 | 92.00% |
| **Online Directories** | 19 | 6 | 76.00% |

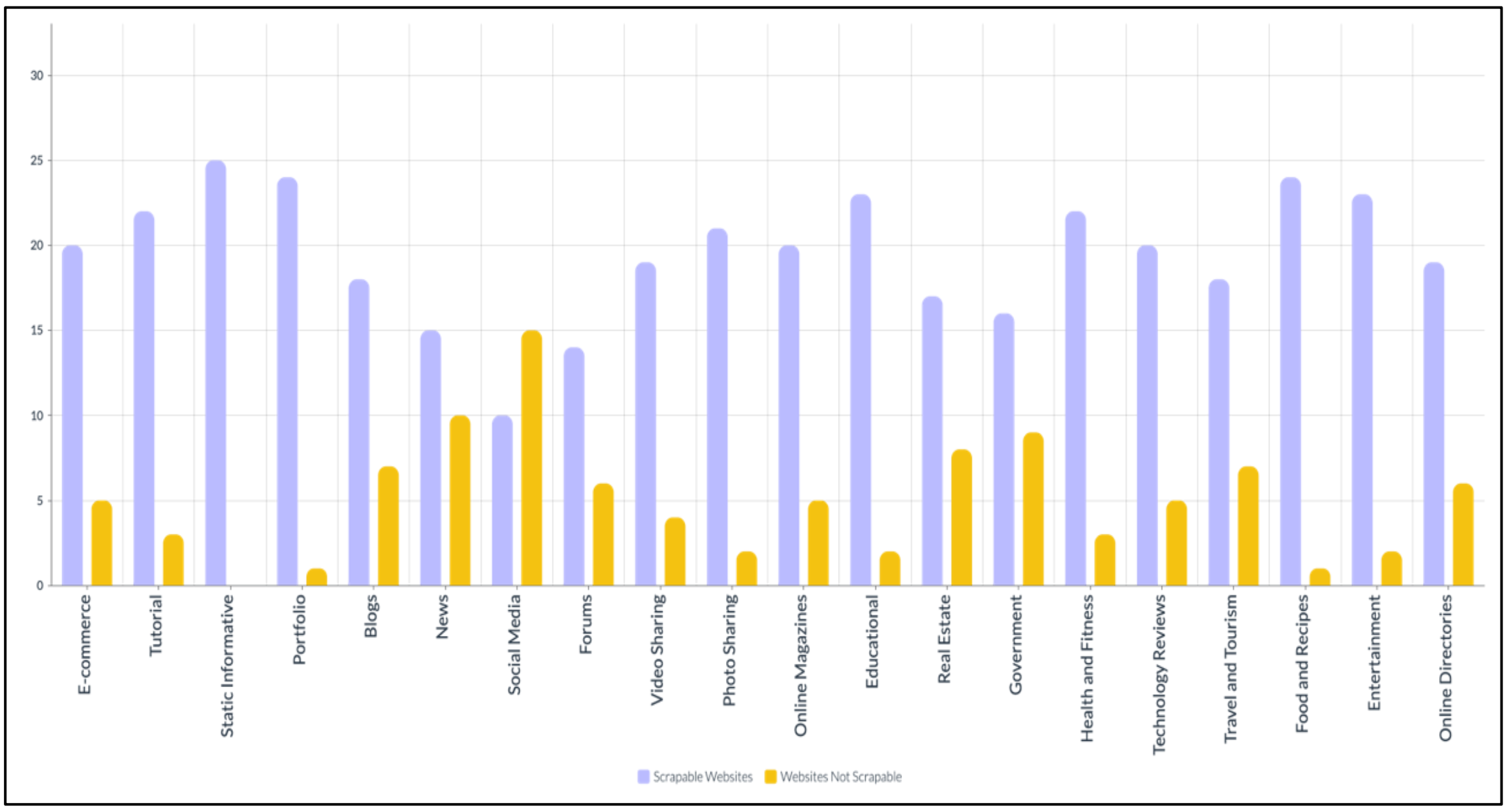

**Fig. 16.** Graphical representation of websites that are scrapable and those which are not scrapable out of sample 100 websites

**Table 2** presents a breakdown of website scrapability by category. The overall scrapability rate across all website categories was moderately high, with an average scrapability rate of **79.40%**. Static informative websites had the highest scrapability rate at **100%**, followed by portfolio websites at **96.00%**, photo sharing websites at **91.30%**, and educational websites at **92.00%**. Other notable categories with high scrapability rates include tutorial websites at **88.00%** and health and fitness websites also at **88.00%**.

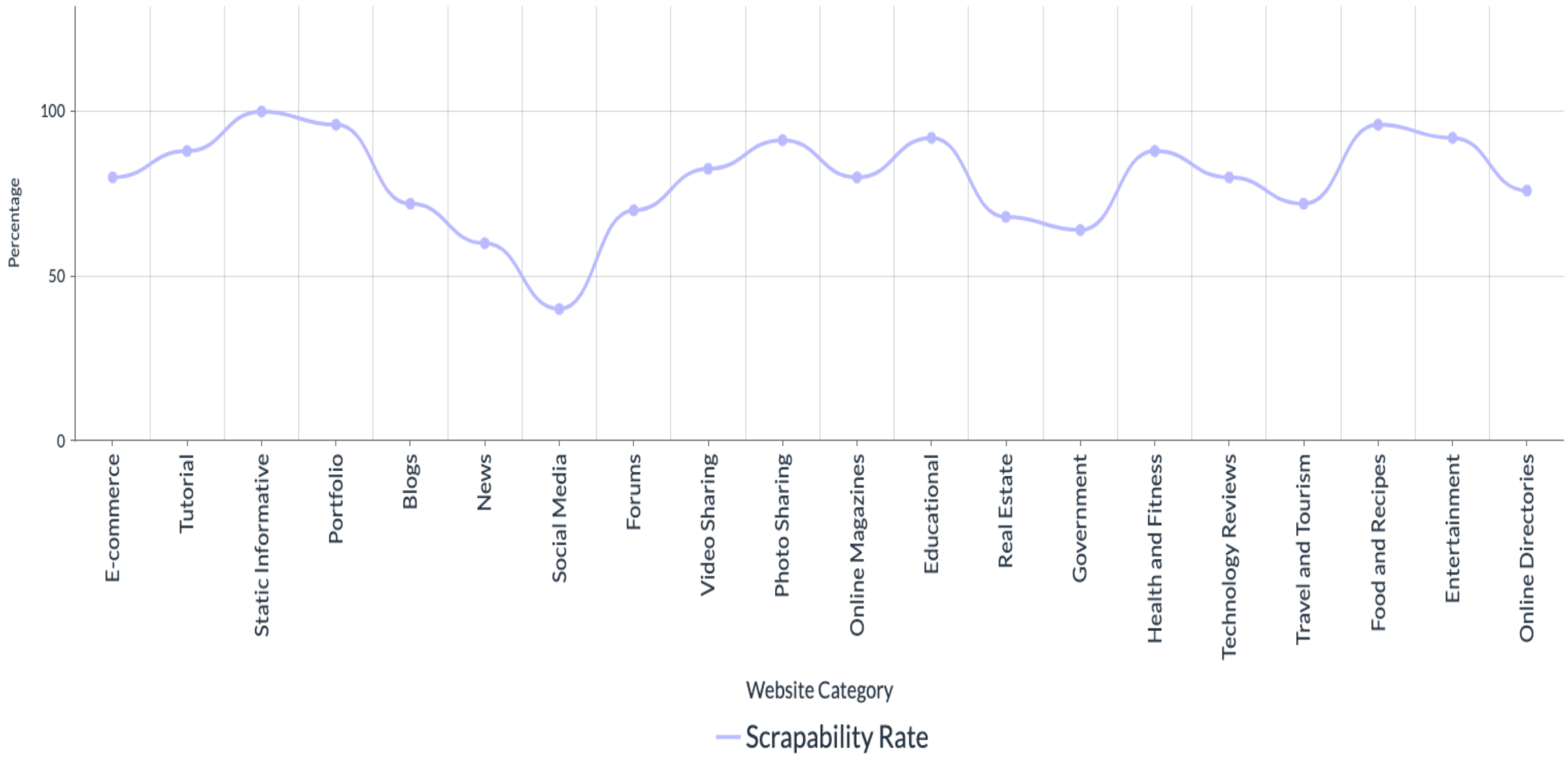

**Fig. 17.** Graphical representation of Scrapability Rate of different categories

On the other end of the spectrum, social media websites had the lowest scrapability rate at **40.00%**, followed by news websites at **60.00%**, and government websites at **64.00%**.

The findings highlight a generally positive trend in website scrapability across various categories, suggesting that many websites are becoming more technically accessible to data extraction tools, potentially due to a focus on structured content and standardized coding practices. However, the average scrapability rate also indicates that a significant portion of websites remains challenging to scrape.

The higher scrapability rate for static informative websites could be attributed to their simpler design and content structure, which contrasts with more complex categories like social media and e-commerce websites that often feature dynamic content and intricate functionalities. Portfolio websites may also emphasize well-structured content to showcase design work effectively, making them more accessible to data scraping tools used for web archiving or design analysis.

Mathematical model of calculation of average runtime and memory usage is given below:

The runtime of the content extraction process depends on the number of nodes $n$ in the HTML DOM tree that are traversed, and the number of relevant nodes $m$ from which data is extracted. The total runtime $T(n)$ can be expressed as:

$$T(n) = T_{\text{traversal}}(n) + T_{\text{extraction}}(m)$$

Where:

- $T_{\text{traversal}}(n)$ is the time required to traverse the DOM structure with $n$ nodes.
- $T_{\text{extraction}}(m)$ is the time required to extract data from $m$ relevant nodes (with $m \leq n$).

The traversal time $T_{\text{traversal}}(n)$ is proportional to the number of nodes and can be modeled as:

$$T_{\text{traversal}}(n) = c_1 \cdot n$$

where $c_1$ is a constant representing the average time taken per node traversal.

The extraction time $T_{\text{extraction}}(m)$ is proportional to the number of relevant nodes and can be modeled as:

$$T_{\text{extraction}}(m) = c_2 \cdot m$$

where $c_2$ is the average time taken to extract data from each relevant node.

Thus, the total runtime $T(n)$ is given by:

$$T(n) = c_1 \cdot n + c_2 \cdot m$$

The memory usage $M(n)$ depends on the amount of memory required during traversal and the amount of memory used to store the extracted data. The total memory usage can be expressed as:

$$M(n) = M_{\text{traversal}}(n) + M_{\text{extraction}}(m)$$

Where:

- $M_{\text{traversal}}(n)$ is the memory used to store information during traversal.
- $M_{\text{extraction}}(m)$ is the memory used to store the extracted data from the $m$ relevant nodes.

The memory used during traversal is proportional to the number of nodes traversed:

$$M_{\text{traversal}}(n) = c_3 \cdot n$$

where $c_3$ is the average memory used per node.

The memory used for storing extracted data is proportional to the number of relevant nodes:

$$M_{\text{extraction}}(m) = c_4 \cdot m$$

where $c_4$ is the average memory used to store the extracted data from each relevant node.

Thus, the total memory usage $M(n)$ is given by:

$$M(n) = c_3 \cdot n + c_4 \cdot m$$

The final equations for the runtime and memory usage in web content extraction are:

**Runtime Calculation**:

$$T(n) = c_1 \cdot n + c_2 \cdot m$$

**Memory Usage Calculation**:

$$M(n) = c_3 \cdot n + c_4 \cdot m$$

Where:

- $n$ is the total number of nodes traversed.
- $m$ is the number of relevant nodes from which data is extracted.
- $c_1, c_2, c_3, c_4$ are constants representing the time and memory costs for traversing and extracting data.

In this study, we have used the above mentioned mathematical model to calculate the runtime and memory consumption of various web scraping tools and libraries. This approach was chosen over generic time calculation methods because it provided more accurate and consistent results. By applying these mathematical models, we were able to derive precise average runtimes and memory utilization for each tool, as demonstrated in the provided data. For example, using these calculations, lxml was determined to be the fastest with an execution time of 917.67 ms, and Selenium was identified as the most memory-intensive tool with an average consumption of 200 MB. This accuracy is essential for an in-depth comparison and understanding of the performance characteristics of each web scraping method.

The results of these calculations are represented in the tabular and graphical representation given below, offering a clear visualization of the performance metrics across different tools and libraries.

**Table. 3.** Web scrapping libraries and automated web scraper average runtime and memory consumption.

| Tool/Library | Average Runtime Using Mathematical Model (ms) | Average Memory Consumption Using Mathematical Model (MB) | Average Runtime without using Mathematical Model (ms) | Average Memory Consumption without using Mathematical Model (MB) |
|---|---|---|---|---|
| **BeautifulSoup** | 1899.00 | 47 | 1897.00 | 45 |
| **Selenium** | 15397.33 | 200 | 15395.33 | 197 |
| **Scrapy** | 1365.67 | 120 | 1363.67 | 117 |
| **MechanicalSoup** | 2019.67 | 60 | 2017.67 | 58 |
| **lxml** | 917.67 | 35 | 915.67 | 33 |
| **Automated Web Scraper** | 6128.66 | 150 | 6126.66 | 148 |

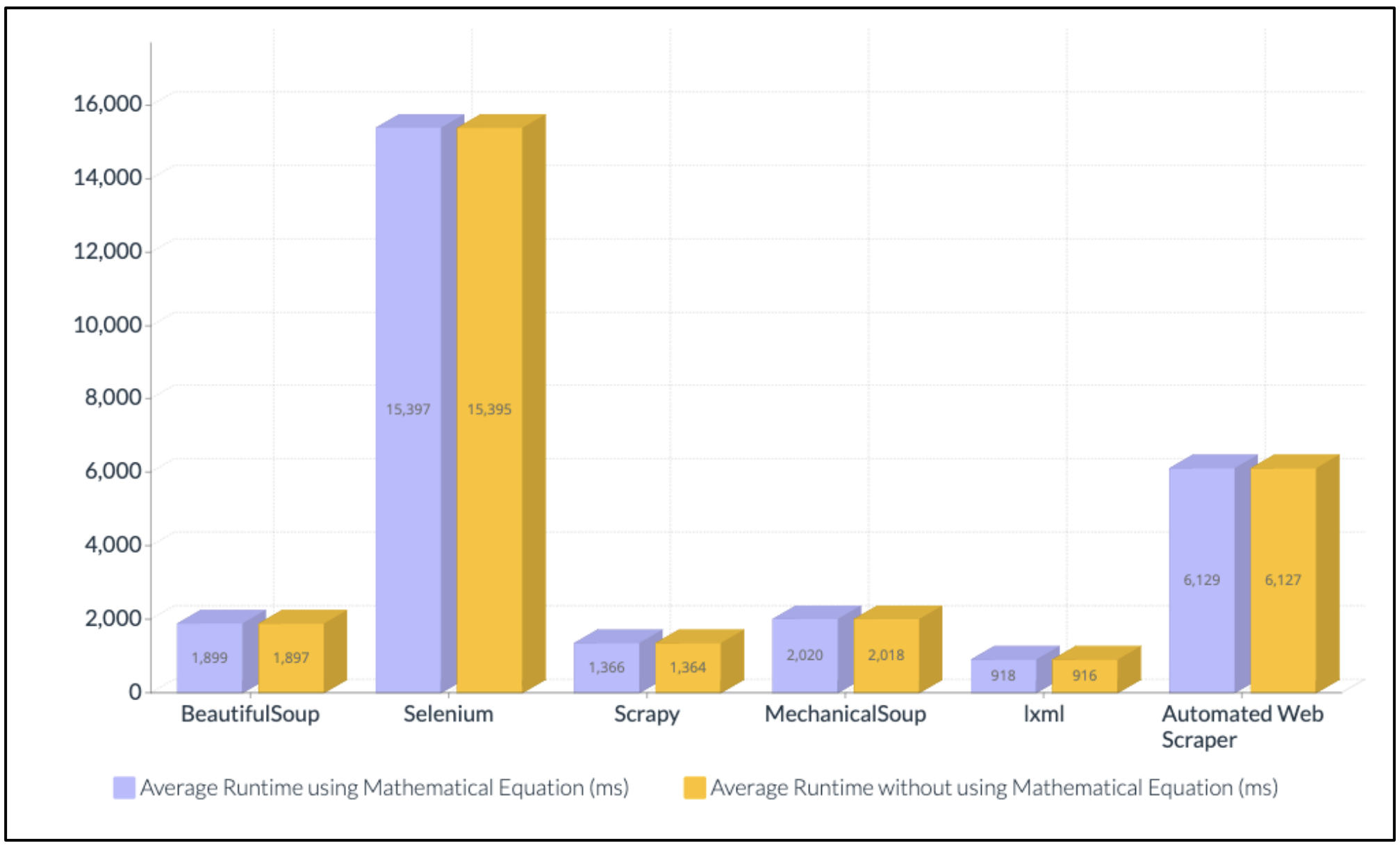

**Fig. 18.** Comparative analysis of Average runtime in milliseconds of various web scrapping libraries and Automated Web Based Web Scrapping Tool using mathematical model and without using mathematical model

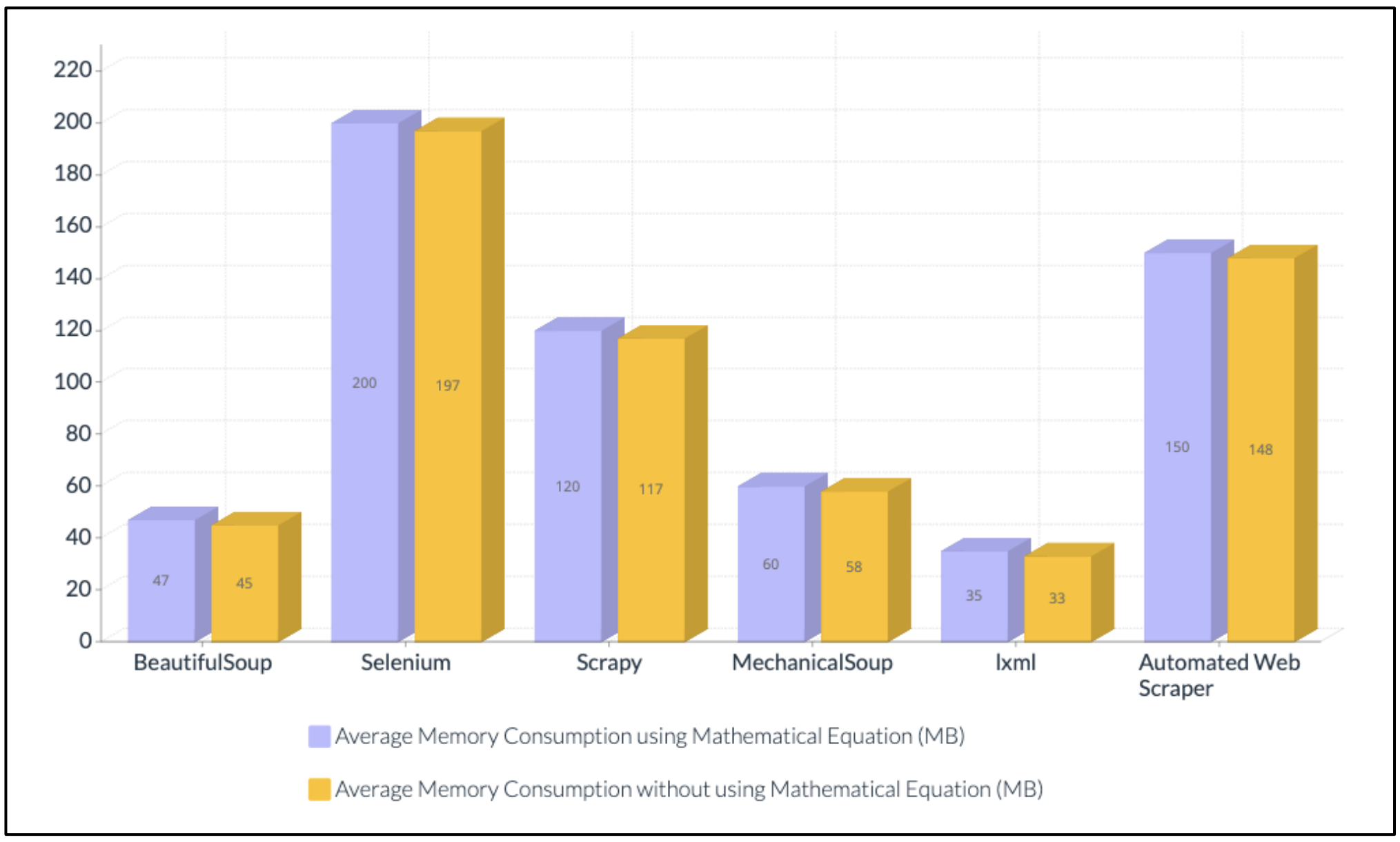

**Fig. 19.** Comparative analysis of Average memory consumption in Megabytes of various web scrapping libraries and Automated Web Based Web Scrapping Tool using mathematical model and without using mathematical model

Overall the performance of the **Automated Web Scraper** is quite competitive when compared to widely used libraries. Despite its higher memory usage and execution time, it offers significant additional functionalities, such as:

1. **Data Processing**: It automatically processes the scraped data and outputs it in CSV format.
2. **Data Persistence**: It maintains a record of scraped data in a **MongoDB database**, which adds to the versatility and usability of the tool in long-term projects.

Thus, while tools like **lxml** and **Scrapy** excel in lightweight scraping tasks with minimal memory and execution time, the **Automated Web Scraper** offers a more robust solution when data persistence and further processing are required.

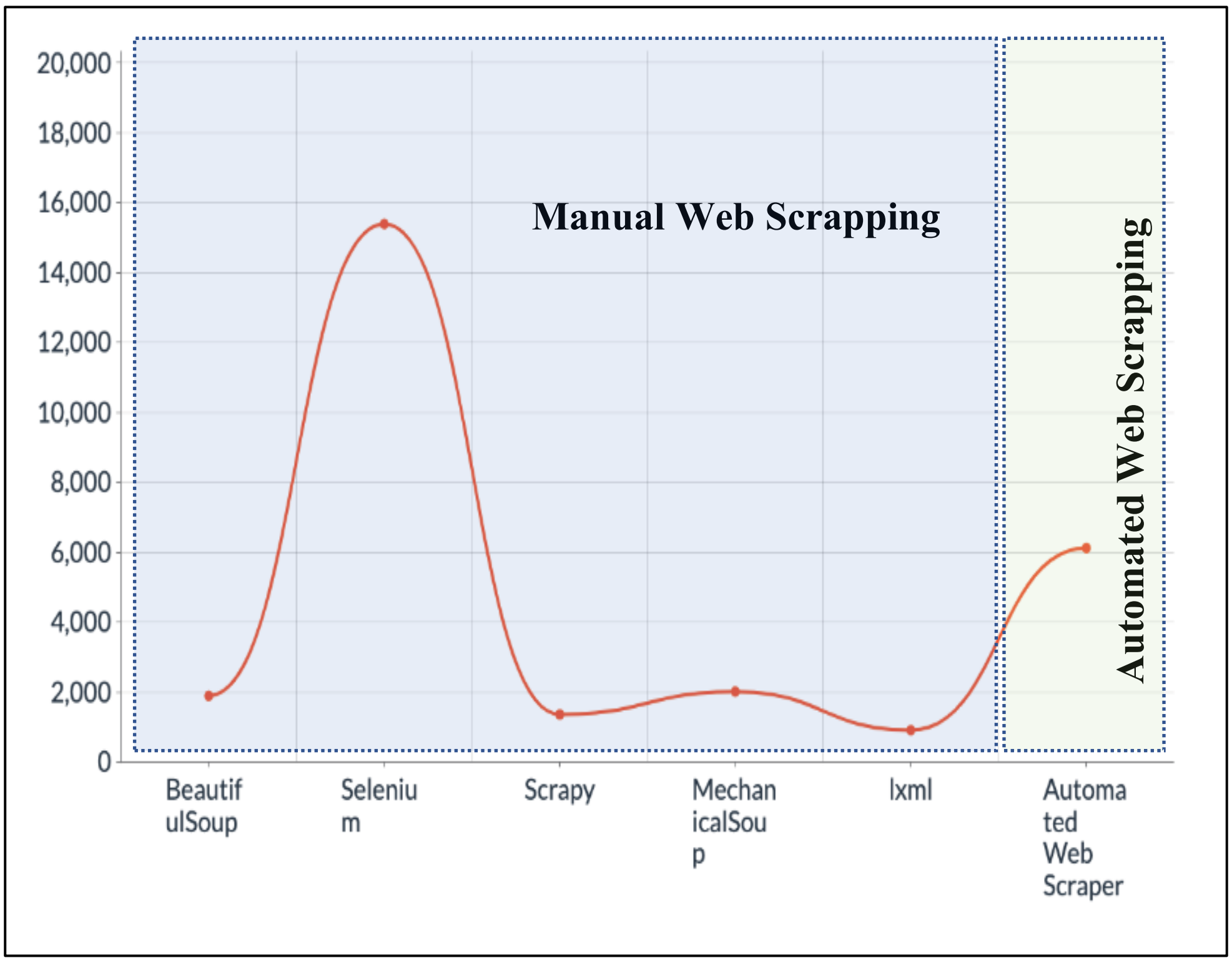

**Fig. 20.** Average runtime in milliseconds of various web scrapping libraries and Automated Web Based Web Scrapping Tool

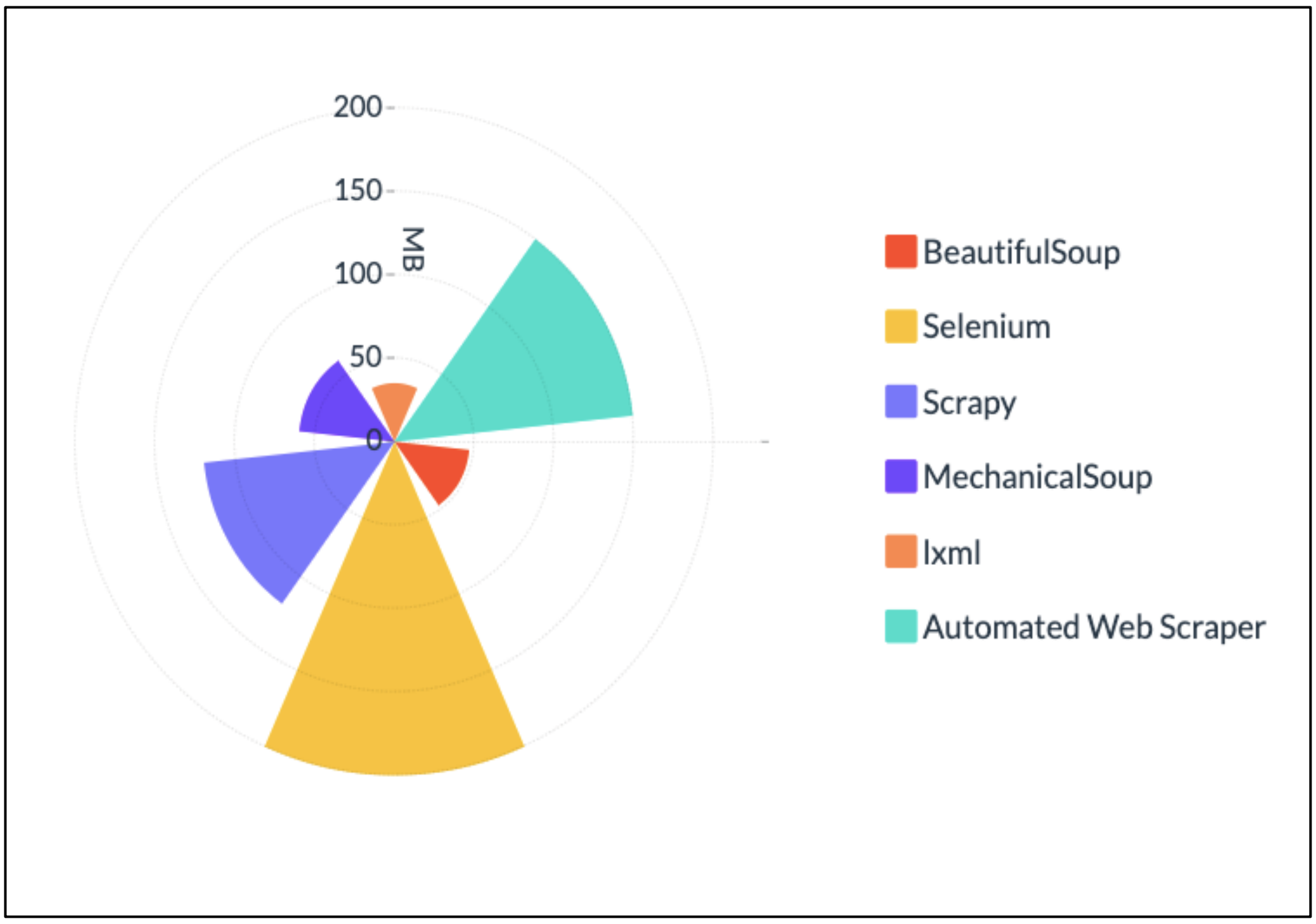

**Fig. 21.** Average memory consumption in Megabytes of various web scrapping libraries and Automated Web Based Web Scrapping Tool

Further investigation is needed to understand the specific challenges faced in scraping categories such as social media, news, and e-commerce websites. This could involve analyzing the technical factors hindering data extraction, such as the use of complex scripts, dynamic content loading, or anti-scraping measures.

## Conclusion

Web scraping has emerged as an essential technique for extracting and utilizing vast amounts of data available on the World Wide Web. This process involves using HTTP or web browsers to convert unstructured data from websites into structured formats that can be stored and analyzed in databases. The utility of web scraping spans various domains, including research, business intelligence, and content aggregation, providing a means to leverage the wealth of information available online. As the internet continues to expand, the relevance and importance of web scraping will only grow, underscoring the need for responsible data handling practices to address concerns related to data privacy and security.

The research presented in this paper demonstrates the successful design and implementation of a user-friendly, automated web application that significantly simplifies the web scraping process for non-technical users. By breaking down the complex task of web scraping into three main stages—fetching, extraction, and execution—the application allows users to efficiently access, extract, and manage data from websites with minimal technical knowledge. The inclusion of user registration and login functionalities, supported by MongoDB, ensures secure and personalized experiences, while the deployment through the Flask framework offers a robust and scalable environment.

The findings from the website scrapability testing, as detailed in the table, highlight the application's broad applicability across various website categories, demonstrating an average scrapability rate of 79.40%. This result underscores the application's ability to handle a diverse range of websites, making it an invaluable tool for users from different domains who require tailored data extraction.

Overall, the methodology and application presented in this research not only optimize the web scraping process but also democratize access to web scraping tools. By making data gathering and management more accessible to a broader audience, this work contributes to the growing need for efficient data utilization in various fields, empowering users to make informed decisions based on the data they can now easily gather and analyze.

## Future Work

To enhance the understanding and application of web scraping, future research should focus on several key areas:

1. Expanding Sample Size: Increasing the sample size of websites used in studies will improve the generalizability of findings, providing more robust insights into the effectiveness and limitations of web scraping across different types of websites.
2. In-depth Technical Analysis: Conducting a more thorough examination of the technical challenges encountered in scraping websites across various categories can help identify specific issues and develop solutions to overcome them. This includes exploring the differences in HTML structure, dynamic content loading, and the use of APIs.

3. Investigating Social Media Websites: Analyzing the reasons behind the lower scrapability rates of social media websites is crucial. This research should focus on understanding the anti-scraping measures employed by these websites, such as CAPTCHAs, rate limiting, and the use of dynamic content. Identifying these measures can help develop more effective scraping techniques and highlight areas for improvement in website design and data accessibility practices.

By addressing these areas, future research can provide a more comprehensive understanding of website scrapability, leading to improved methodologies and best practices for web scraping. This, in turn, will enhance the ability to collect, analyze, and utilize web data responsibly and effectively across various domains.